\documentclass[aps,PRB,twocolumn,superscriptaddress,preprintnumbers]{revtex4-2}
\usepackage[T1]{fontenc}
\usepackage{times}
\usepackage{graphicx,color}
\usepackage{amsfonts,amsmath,amssymb,amsbsy}
\usepackage[colorlinks=true, linkcolor=blue, citecolor=blue, urlcolor=blue]{hyperref}
\usepackage{float}

\begin{document}

\title{Purely electrical detection of the spin-splitting vector in $p$-wave
magnets\\
based on linear and nonlinear conductivities}
\author{Motohiko Ezawa}
\affiliation{Department of Applied Physics, The University of Tokyo, 7-3-1 Hongo, Tokyo
113-8656, Japan}

\begin{abstract}
A $p$-wave magnet has a momentum-dependent spin splitting and zero-net
magnetization just as in the case of an altermagnet. It will be useful for
high-density and ultra-fast memory, where the direction of the
spin-splitting vector may be used as a bit. The spin-splitting vector
corresponds to the N\'{e}el vector in altermagnets. However, it is a
nontrivial problem to detect the spin-splitting vector in a $p$-wave magnet
because time-reversal symmetry is preserved, while this is not a problem in
an altermagnet because the anomalous Hall conductivity is present due to the
breaking of time-reversal symmetry. Here, we show that it is possible to
detect the in-plane component of the spin-splitting vector in the $p$-wave
magnet by measuring the linear transverse and longitudinal Drude
conductivity. Remarkably, this measurement is possible without using
magnetization. Furthermore, we study the nonlinear Drude conductivity, the
quantum-metric and the Berry curvature dipole induced nonlinear conductivity
in the presence of tiny magnetization along the $z$\ axis. It is possible to
detect the $z$-component of the spin-splitting vector by measuring the above
nonlinear conductivities. We obtain analytic formulae for them in the
first-order perturbation theory, which agree quite well with numerical
results without perturbation.
\end{abstract}

\date{\today }
\maketitle

\section{Introduction}

Ferromagnets are useful for nonvolatile memories, where the up and down
spins act as a bit. On the other hand, antiferromagnets are expected to act
as more efficient memories owing to their zero net magnetization, where
high-density and quick-switchable memories may be possible. However, it is
very hard to readout the direction of the N\'{e}el vector of antiferromagnet%
\cite{Jung,Baltz,Han,Ni,Godin,Kimura,ZhangNeel}. Altermagnets are attractive 
\cite{SmejRev,SmejX,SmejX2} because they have both merits of ferromagnets
and antiferromagnets. Altermagnets have zero net magnetization and they
break time-reversal symmetry. Accordingly, the N\'{e}el vector is observable
by means of the anomalous Hall conductivity\cite{Fak,Tsch,Sato,Leiv}, and
hence its direction can be used as a bit. A characteristic feature of
altermagnets which is absent in both ferromagnets and antiferromagnets is
the momentum-dependent spin splitting\cite%
{Ahn,Hayami,Naka,SmejRev,SmejX,SmejX2}, which includes the $d$-wave, $g$%
-wave and $i$-wave band structures. Indeed, momentum dependent band
structures have been observed by Angle-Resolved Photo-Emission Spectroscopy
(ARPES)\cite{Krem,Lee,Fed,Osumi,Lin}. Furthermore, spin current is generated
in $d$-wave altermagnets\cite{Naka,Gonza,NakaB,Bose}.

In general, we may think of the higher symmetric $X$-wave magnets\cite{GI},
where $X=p,d,f,g,i$. They have zero net magnetization as in the case of
antiferromagnets. The spin texture is collinear and the band structure
breaks time reversal symmetry for $X=d,g,i$. They are altermagnets. On the
other hand, the spin texture is noncollinear and the band structure
preserves time reversal symmetry for $X=p,f$, which leads to the zero
anomalous Hall conductivity.\ The effective Hamiltonian\cite{GI} for the $X$%
-wave magnet is concisely described by a two-band model containing the term $%
\left( \mathbf{s}\cdot \mathbf{\sigma }\right) f\left( \mathbf{k}\right) $,
where $s$\ is a three-dimensional constant unit vector representing the
spin-splitting direction, and $f\left( \mathbf{k}\right) $\ is a symmetric
function characterizing the $X$-wave magnet.\ We may call $s$\ the
spin-splitting vector. It is identical to the N\'{e}el vector in
altermagnets.

Recently, $p$-wave magnets attract growing interests\cite%
{pwave,Martin,H20,H20b,H22,Okumura,Kuda,Brek,EzawaPwave,GI,Chak,Edel}. The $p
$-wave symmetry is concisely described by the term $\left( \mathbf{s}\cdot 
\mathbf{\sigma }\right) k_{x}$\ in the Hamiltonian. This term is actually
derived from the four-band Hamiltonian\cite{pwave,Chak} constructed for the $%
p$-wave magnet CeNiAsO\ based on the density-functional theory, as we show
in this work. The direction of the spin-splitting vector can be used as a
bit. However, there is so far no known method to detect the spin-splitting
vector, which makes difficult to use it for spintronics. Experiments on the $%
p$-wave magnets\ were reported\cite{Yamada} in Gd$_{3}$Ru$_{4}$Al$_{12}$\
and reported\cite{Comin} in NiI$_{2}$. Moreover, an experiment on an
electrical switch of the spin-splitting vector was reported\cite{Comin} in
the $p$-wave magnet NiI$_{2}$.

The linear conductivity $\sigma ^{a;b}$ is defined by $j^{a;b}=\sigma
^{a;b}E^{a}$, where $E^{a}$ is an applied electric field along the $a$
direction and $j^{a;b}$ is the current along the $b$ direction. They include
the longitudinal Drude conductivity $\sigma ^{x;x}$ and the transverse Drude
conductivity $\sigma ^{y;x}$. On the other hand, there are several studies
on nonlinear conductivity\cite%
{Gao,Sodeman,HLiu,Michishita,Watanabe,CWang,Oiwa,AGao,NWang,KamalDas,Kaplan,Ohmic,Xiang}%
. Especially, the second-order nonlinear conductivity $\sigma ^{ab;c}$ is
defined by $j^{ab;c}=\sigma ^{ab;c}E^{a}E^{b}$, where $E^{a}$ is an applied
electric field along the $a$ direction and $j^{ab;c}$ is the current along
the $c$ direction. Among them, the nonlinear Drude conductivity is
proportional to $\tau $, where $\tau $ is the electron relaxation time. It
is an extrinsic nonlinear conductivity, while the quantum-metric induced
nonlinear conductivity is intrinsic because it is irrelevant to $\tau $.

In this paper, we show that the in-plane component of the spin-splitting
vector of $p$-wave magnets is detectable by measuring the transverse and
longitudinal linear Drude conductivities. Remarkably, this is possible
without using magnetization. In addition, the $z$-component of the
spin-splitting vector of $p$-wave magnets is detectable by measuring the
nonlinear conductivity with the aid of the magnetization along the $z$ axis.

This paper is composed as follows. In Sec.\ref{Model}, we introduce the
minimal two-band Hamiltonian describing $p$-wave magnets. We also derive
this minimal Hamiltonian from the four-band Hamiltonian\cite{pwave,Chak}
constructed based on the density-functional theory. In Sec.\ref{Linear}, we
study linear conductivities including the Drude conductivity and the
anomalous Hall conductivity. We show that the in-plane component of the
spin-splitting vector is observable without applying magnetic field. In Sec.%
\ref{Nonlinear}, we study nonlinear conductivities including the nonlinear
Drude conductivity, the quantum-metric induced nonlinear conductivity and
the Berry curvature dipole induced nonlinear conductivity. The out-of-plane
component of the spin-splitting vector is observable in the presence of
magnetic field. Sec.\ref{Discussion} is devoted to discussions.

\section{Model\label{Model}}

\subsection{Two-band Hamiltonian}

\begin{figure}[t]
\centerline{\includegraphics[width=0.48\textwidth]{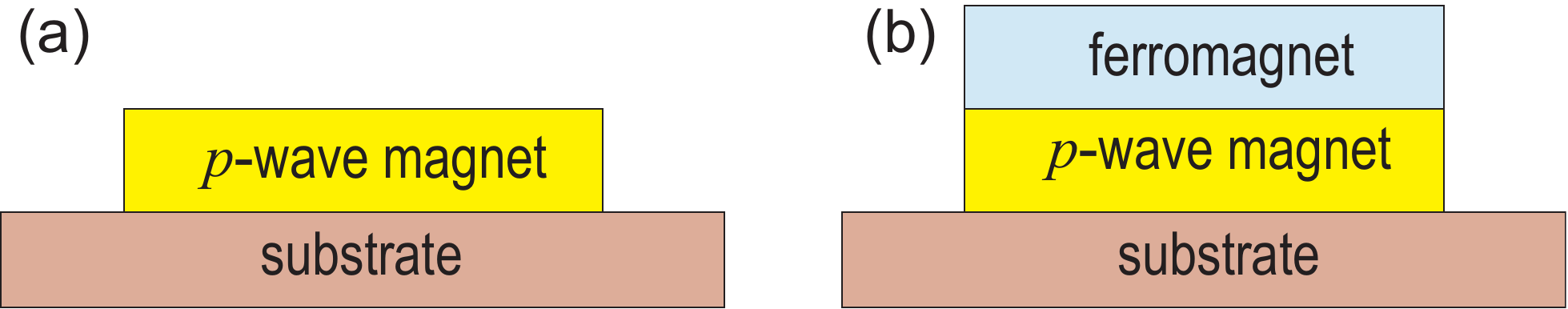}}
\caption{Illustration of the system made of (a) $p$-wave-magnet-substrate
structure and (b) ferromagnet-$p$-wave-magnet-substrate structure.}
\label{FigSub}
\end{figure}

We analyze a system where a $p$-wave magnet is placed on a substrate as
illustrated in Fig.\ref{FigSub}(a). The substrate breaks inversion symmetry
and induces the Rashba interaction. Then, the minimal two-band Hamiltonian
is given by%
\begin{equation}
H=H_{0}+H_{\lambda }+H_{p}+H_{B}.  \label{TotalHamil}
\end{equation}%
The first term represents the kinetic energy of free electrons present in
the interface,%
\begin{equation}
H_{0}=\frac{\hbar ^{2}\left( k_{x}^{2}+k_{y}^{2}\right) }{2m}\sigma _{0},
\end{equation}%
where $m$ is the free-fermion mass, and $\sigma _{0}$ is $2\times 2$
identity matrix.

The second term represents the Rashba interaction,%
\begin{equation}
H_{\lambda }=\lambda \left( -k_{y}\sigma _{x}+k_{x}\sigma _{y}\right) ,
\label{BHZ}
\end{equation}%
where $\lambda $ is the magnitude of the spin-orbit interaction, and $\sigma
_{x}$ and $\sigma _{y}$ are the Pauli matrices for the spin. It is
introduced by making an interface between the $p$-wave magnet and the
substrate breaking inversion symmetry.

The third term%
\begin{equation}
H_{p}\left( \mathbf{k}\right) =J\left( \mathbf{s}\cdot \mathbf{\sigma }%
\right) k_{x}  \label{HJ}
\end{equation}%
describes the effect of the $p$-wave magnet on electrons in the interface%
\cite{pwave,Maeda,EzawaPwave,Brek,EzawaPwave,GI,Edel,Edel}, where $J$\ is
the $s$-$d$\ coupling constant between the localized spin of the $p$-wave
magnet and free electrons in the interface, and $s$\ is the
three-dimensional constant unit vector parametrized as%
\begin{equation}
\mathbf{s}=\left( \sin \Theta \cos \Phi ,\sin \Theta \sin \Phi ,\cos \Theta
\right) 
\end{equation}%
with constants $\Theta $ and $\Phi $. It is called the spin-splitting
vector. We set 
\begin{equation}
J_{x}=J\sin \Theta \cos \Phi ,\quad J_{y}=J\sin \Theta \sin \Phi ,\quad
J_{z}=J\cos \Theta .
\end{equation}%
A comment is in order on the spin-splitting vector. When the noncollinear
helical spin texture leads to the $p$-wave spin-splitting band structure,
the spin-splitting vector $s$\ is orthogonal to the helix rotation
direction. Indeed, the generator of the helix is given by%
\begin{equation}
\exp \left[ i\left( \mathbf{s}\cdot \mathbf{\sigma }\right) k_{x}x\right] ,
\end{equation}%
where the rotation angle of the helix is orthogonal to the spin-splitting
vector. For example, if the spin rotates in the $x$-$y$\ plane, the
spin-splitting vector is along the $z$\ axis.

\begin{figure}[t]
\centerline{\includegraphics[width=0.48\textwidth]{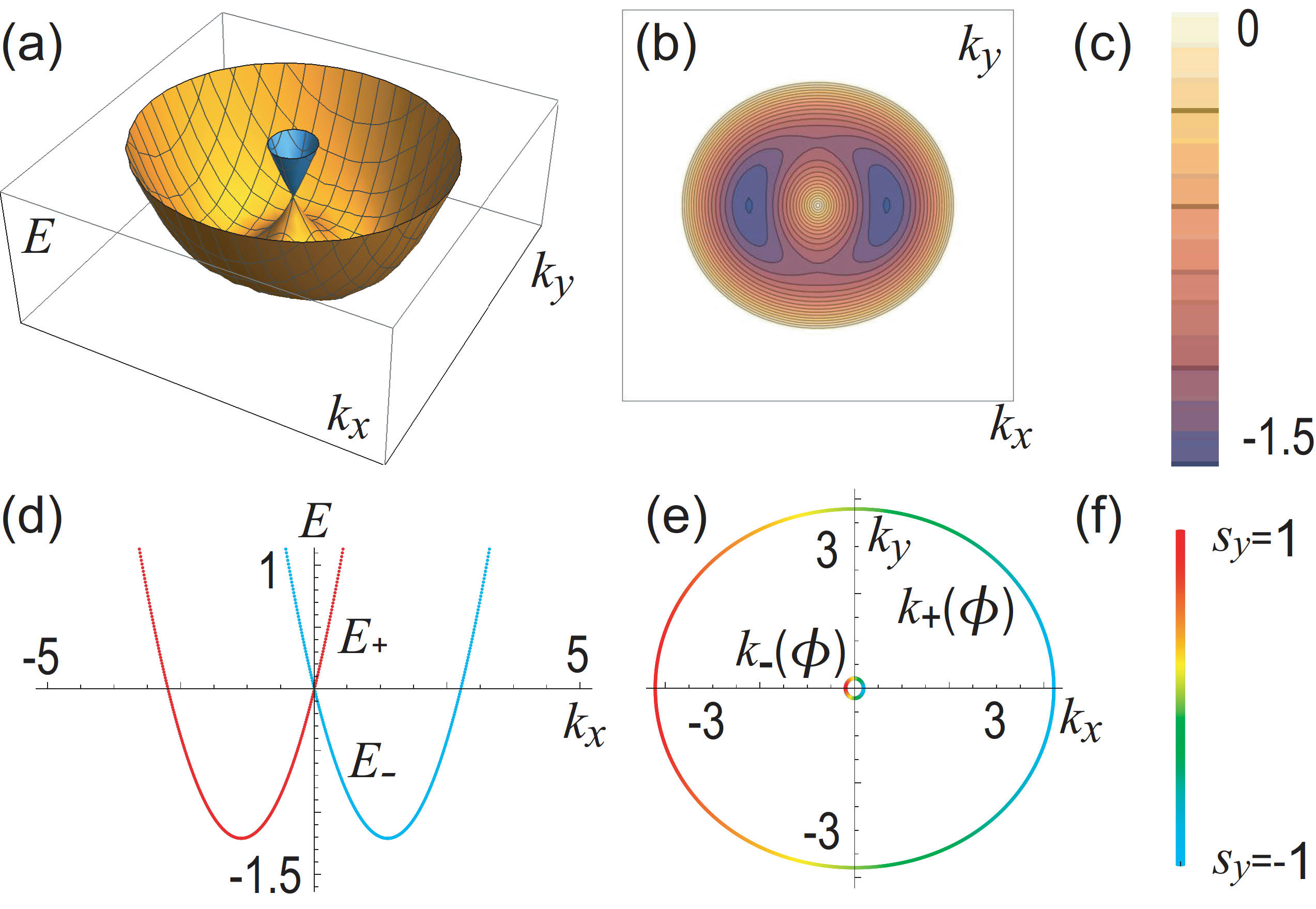}}
\caption{(a) Bird's eye's view of the band structure, where the
spin-splitting vector is taken along the $y$ axis with $\Theta =\protect\pi %
/2$\ and $\Phi =\protect\pi /2$. (b) Its contour plot with the color palette
(c). (d) Its cross section along the $k_{x}$ axis. The horizontal axis is $%
k_{x}$ in units of $k_{0}$. (e) Fermi surfaces at $\protect\mu =-0.2\protect%
\varepsilon _{0}$. Two ellipses represent $p$-wave symmetric Fermi surfaces $%
k_{\pm }\left( \protect\phi \right) $, with $k_{x}=k\cos \protect\phi $ and $%
k_{y}=k\sin \protect\phi $. (f) The color palette for (d) and (e), where $%
s_{y}$\ is shown in the color. We have set $\hbar ^{2}k_{0}^{2}/\left(
2m\right) =\protect\varepsilon _{0}/2$, $\protect\lambda =\protect%
\varepsilon _{0}/k_{0}$ and $J=0.5\protect\varepsilon _{0}/k_{0}$ with $%
\protect\varepsilon _{0}\equiv m\protect\lambda ^{2}/2\hbar ^{2}$ and $%
k_{0}=m\protect\lambda /2\hbar ^{2}$. }
\label{FigBandY}
\end{figure}

The last term%
\begin{equation}
H_{B}=B\sigma _{z}  \label{Zeeman}
\end{equation}%
describes the Zeeman term, which breaks time-reversal symmetry. This term
may be introduced either by attaching a ferromagnet as illustrated in Fig.%
\ref{FigSub}(b), or by applying external magnetic field. The breaking of
time-reversal symmetry is necessary for nonzero nonlinear conductivities\cite%
{Kaplan}.

The energy spectrum of the Hamiltonian (\ref{TotalHamil}) is given by%
\begin{equation}
\varepsilon _{\pm }=\frac{\hbar ^{2}k^{2}}{2m}\pm \sqrt{%
h_{x}^{2}+h_{y}^{2}+h_{z}^{2}},  \label{Epm}
\end{equation}%
where%
\begin{align}
h_{x}=& -\lambda k_{y}+Jk_{x}\sin \Theta \cos \Phi ,  \notag \\
h_{y}=& \lambda k_{x}+Jk_{x}\sin \Theta \sin \Phi ,  \notag \\
h_{z}=& B+Jk_{x}\cos \Theta .
\end{align}%
We introduce the polar coordinate $k_{x}=k\cos \phi $ and $k_{y}=k\sin \phi $%
. In the first order of $J$ and $B$, the energy is expanded as%
\begin{align}
\varepsilon _{\pm }& =\frac{\hbar ^{2}k^{2}}{2m}  \notag \\
& \pm \left( \frac{BJ}{\lambda }\cos \Theta \cos \phi +k\lambda -Jk\cos \phi
\sin \Theta \sin \left( \phi -\Phi \right) \right) .
\end{align}

The band structure is shown in Fig.\ref{FigBandY}, where the spin-splitting
vector is taken along the $x$\ axis. A Dirac cone exists at the momentum $%
k_{x}=k_{y}=0$ as shown in Fig.\ref{FigBandY}(a) and (c), which is formed by
the Rashba interaction. We set $\bar{m}\equiv m/\hbar ^{2}$\ in the
following.

There are two Fermi surfaces characterized by $\eta =\pm $ for the energy $%
\varepsilon _{\chi }$ with $\chi =\pm $, which are obtained as%
\begin{align}
k_{\eta }^{\chi }\left( \phi \right) =& -\chi \bar{m}\lambda +\eta \bar{m}%
J\cos \phi \sin \Theta \sin \left( \phi -\Phi \right)  \notag \\
& +\eta (\frac{2\left( \mu \lambda -\chi BJ\cos \Theta \cos \phi \right) }{%
\bar{m}\lambda }  \notag \\
& +\left( \lambda -J\cos \phi \sin \Theta \sin \left( \phi -\Phi \right)
\right) ^{2})^{1/2},  \label{kpm}
\end{align}%
where $\eta =1$\ represents the outer Fermi surface and $\eta =-1$\
represents the inner Fermi surface. The Fermi surfaces have the $p$-wave
symmetry as shown in Fig.\ref{FigBandY}(b). There are two Fermi surfaces $%
k_{\pm }\left( \phi \right) $ as shown in Fig.\ref{FigBandY}(d), where $%
k_{x}=k\cos \phi $ and $k_{y}=k\sin \phi $.

\subsection{Derivation of the two-band Hamiltonian\label{4to2}}

\begin{figure}[t]
\centerline{\includegraphics[width=0.48\textwidth]{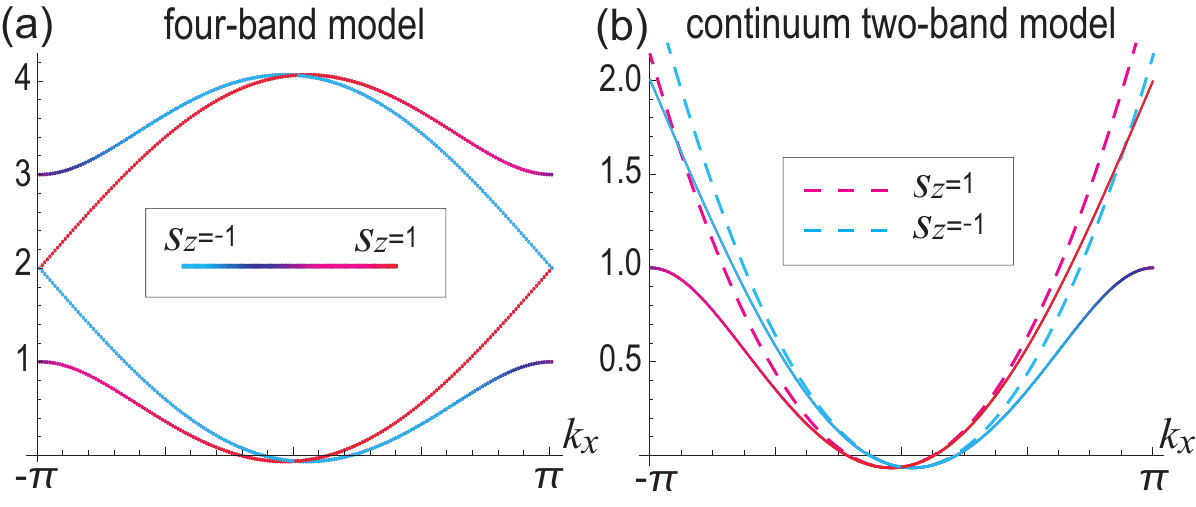}}
\caption{(a) The band structure of the four-band model (\protect\ref{CNAO}).
The color shows the $s_{z}$\ component, where the color palette is shown in
the inset. (b) Comparison of the band structures of the lowest two bands
between the four-band model (\protect\ref{CNAO}) and the two-band continuum
model (\protect\ref{Heff2}). The solid curves are based on the four-band
Hamiltonian (\protect\ref{CNAO}). Dashed curves are based on the two-band
continuum Hamiltonian (\protect\ref{Heff2}), where $s_{z}=1$\ is colored in
magenta and $s_{z}=-1$\ is colored in cyan. We have set $t=1$ and $t_{J}=0.25
$.}
\label{FigCNAO}
\end{figure}

We discuss the relation between the minimal two-band Hamiltonian used in
this work and the four-band Hamiltonian\cite{pwave,Chak} derived for the $p$%
-wave magnet CeNiAsO\ based on the density-functional theory. The four-band
Hamiltonian reads 
\begin{eqnarray}
H &=&2t\left( \tau _{x}\cos \frac{ak_{x}}{2}+\cos ak_{y}\right)   \notag \\
&&+2t_{J}\left( \sigma _{x}\tau _{y}\sin \frac{ak_{x}}{2}+\sigma _{y}\tau
_{z}\cos ak_{y}\right) ,  \label{CNAO}
\end{eqnarray}%
where $\sigma $\ and $\tau $\ correspond to the spin and site degrees of
freedom, $t$\ is the hopping amplitude with $t=0.8$eV, $t_{J}$\ is the
magnitude of the $p$-wave magnet with $t_{J}=0.2$eV, and $a$\ is the lattice
constant $a=8.1$\r{A}. The band structure is shown in Fig.\ref{FigCNAO}(a).

We first diagonalize the Hamiltonian (\ref{CNAO}) at the $\Gamma $ point,%
\begin{equation}
U_{0}H\left( k_{x}=0,k_{y}=0\right) U_{0}^{-1}=\text{diag.}\left\{
E_{-},E_{-},E_{+},E_{+}\right\} ,  \label{diago}
\end{equation}%
where the eigen-energies are%
\begin{equation}
E_{\pm }=2\left( t\pm \sqrt{t^{2}+t_{J}^{2}}\right) 
\end{equation}%
with two-fold degeneracy. The eigenvectors are given by%
\begin{eqnarray}
\psi _{\pm }^{a} &=&c_{a}\left\{ \pm \frac{it_{J}}{t},\pm \frac{\sqrt{%
t^{2}+t_{J}^{2}}}{t},0,1\right\} ^{t}, \\
\psi _{\pm }^{b} &=&c_{b}\left\{ \pm \frac{\sqrt{t^{2}+t_{J}^{2}}}{t},\pm 
\frac{it_{J}}{t},1,0\right\} ^{t},
\end{eqnarray}%
where $c_{a}$\ and $c_{b}$\ are normalization constants. The unitary
transformation $U_{0}$\ in Eq.(\ref{diago}) is given by \ 
\begin{equation}
U_{0}=\left\{ \psi _{-}^{a},\psi _{-}^{b},\psi _{+}^{a},\psi
_{+}^{b}\right\} 
\end{equation}%
with the use of the eigenvectors.\ 

\begin{figure}[t]
\centerline{\includegraphics[width=0.48\textwidth]{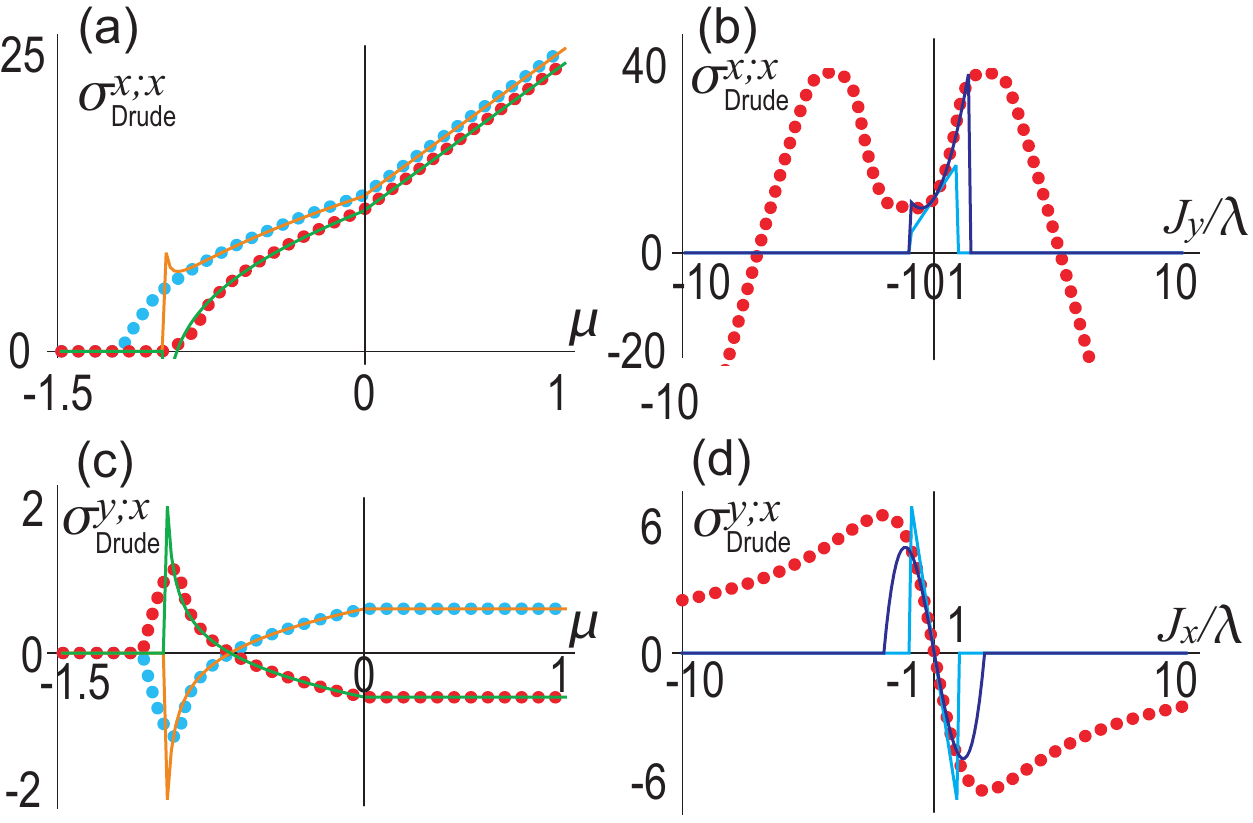}}
\caption{(a) and (b) Longitudinal Drude conductivity $\protect\sigma _{\text{%
Drude}}^{x;x}$ in units of $e^{2}\protect\tau /\hbar ^{2}$, where the
spin-splitting vector is taken along the $y$ axis. (c) and (d) Transverse
Drude conductivity $\protect\sigma _{\text{Drude}}^{y;x}$ in the unit of $%
e^{2}\protect\tau /\hbar ^{2}$, where the spin-splitting vector is taken
along the $x$\ axis. (a) and (c) The horizontal axis is the chemical
potential $\protect\mu $ in units of $\protect\varepsilon _{0}$. (b) The
horizontal axis is $J_{y}/\protect\lambda $. (d) The horizontal axis is $%
J_{x}/\protect\lambda $. (a) and (c) Red (cyan) dots indicate numerically
obtained results by setting $J=0.1\protect\varepsilon _{0}/k_{0}$ ($J=-0.1%
\protect\varepsilon _{0}/k_{0}$). Green (orange) curves indicate
analytically obtained results up to linear order in $J/\protect\lambda $ by
setting $J=0.1\protect\varepsilon _{0}/k_{0}$ ($J=-0.1\protect\varepsilon %
_{0}/k_{0}$). (b) and (d) Red dots (cyan lines) indicate numerically
(analytically up to linear order in $J/\protect\lambda $) obtained results
by setting $\protect\mu =-0.2\protect\varepsilon _{0}$. Blue curves
represent analytic results up to the second order in $J/\protect\lambda $\
in (b) and up to the third order in $J/\protect\lambda $ in (d). We have set 
$\hbar ^{2}k_{0}^{2}/\left( 2m\right) =\protect\varepsilon _{0}/2$ and $%
\protect\lambda =\protect\varepsilon _{0}/k_{0}$.}
\label{FigDrude}
\end{figure}

Then, we make a unitary transformation with the use of the matrix $%
U_{0}$\ in the vicinity of the $\Gamma $\ point, 
\begin{equation}
U_{0}H\left( k_{x},k_{y}\right) U_{0}^{-1}.
\end{equation}%
Taking the lowest two bands, we obtain%
\begin{eqnarray}
H_{\text{eff}} &=&-2\left\vert t\right\vert \cos \frac{ak_{x}}{2}+\frac{%
t_{J}^{2}}{\left\vert t\right\vert }\left( \cos \frac{ak_{x}}{2}-2\cos
ak_{y}\right)   \notag \\
&&+\frac{2t_{J}^{2}}{t}\sin \frac{ak_{x}}{2}\sigma _{z},  \label{HeffTight}
\end{eqnarray}%
up to the second order in $J$. It is expanded in $k_{x}$\ and $k_{y}$\ as%
\begin{eqnarray}
H_{\text{eff}} &=&-2\left\vert t\right\vert \left( 1-\frac{\left(
ak_{x}\right) ^{2}}{8}\right)   \notag \\
&&+\frac{t_{J}^{2}}{\left\vert t\right\vert }\left( \left( 1-\frac{\left(
ak_{x}\right) ^{2}}{8}\right) -2\left( 1-\frac{\left( ak_{y}\right) ^{2}}{2}%
\right) \right)   \notag \\
&&+\frac{t_{J}^{2}}{t}ak_{x}\sigma _{z},
\end{eqnarray}%
which is summarized as%
\begin{equation}
H_{\text{eff}}=\frac{\hbar ^{2}k_{x}^{2}}{2m_{x}}+\frac{\hbar ^{2}k_{y}^{2}}{%
2m_{y}}+Jk_{x}\sigma _{z}+\text{const},  \label{Heff2}
\end{equation}%
with%
\begin{eqnarray}
\frac{\hbar ^{2}}{2m_{x}} &\equiv &\frac{a^{2}}{4}+\frac{t_{J}^{2}a^{2}}{%
4\left\vert t\right\vert },\qquad \frac{\hbar ^{2}}{2m_{y}}\equiv \frac{%
t_{J}^{2}a^{2}}{\left\vert t\right\vert }, \\
J &\equiv &\frac{t_{J}^{2}}{t}a.
\end{eqnarray}%
It is essentially identical to the term (\ref{HJ}) of the minimal two-band
Hamiltonian with $s=\left( 0,0,1\right) $\ except for the mass anisotropy
for $m_{x}\neq m_{y}$. The mass anisotropy does not play any role in the $p$%
-wave magnet by scaling $k_{y}/\sqrt{m_{y}}\mapsto k_{y}^{\prime }/\sqrt{%
m_{x}}$. Note that there is no Rashba interaction in this model. The
two-band model (\ref{Heff2}) reproduces well the lowest two bands of the
four-band model (\ref{CNAO}) in a wide range of $k_{x}$\ around the $\Gamma $%
\ point as shown in Fig.\ref{FigCNAO}(b).\ 

\subsection{Material parameters}

We summarize material parameters in the Hamiltonian (\ref{TotalHamil}). The
mass is given by $m=\hbar ^{2}/\left( ta^{2}\right) $, where it is estimated%
\cite{pwave,Chak,Edel} that the hopping amplitude $t=1$eV and the lattice
constant $a=8.1$\AA . The magnitude of the band splitting is estimated\cite%
{pwave,Chak} as 200meV. The strength of the Rashba interaction is $\lambda
=0.03$eV\AA\ at the Ag(111) surface\cite{Cer}, $\lambda =0.07$eV\AA\ at the
InGaAs/InAlAs surface\cite{Nitta}, and $\lambda =0.33$eV\AA\ at the Au(111)
surface\cite{La}. A typical momentum is $k=\bar{m}\lambda $ from Eq.(\ref%
{kpm}). We may argue that the energy of this Hamiltonian is estimated as $%
\bar{m}\lambda ^{2}$, which is of the order of $1$meV by using $\lambda
=0.33 $eV\AA .

\section{Linear conductivity}

\label{Linear}

\subsection{Drude conductivity}

We first study linear conductivities. We set $B=0$\ in the Hamiltonian (\ref%
{Zeeman}) in this subsection. The Drude conductivity is given by the formula%
\cite{YFang}%
\begin{equation}
\sigma ^{a;b}=\frac{e^{2}\tau }{\hbar ^{2}}\sum_{n}\int d^{2}{k}f_{n}\frac{%
\partial ^{2}\varepsilon _{n}}{\partial k_{a}\partial k_{b}},  \label{Drude}
\end{equation}%
where $f_{n}=1/\left( \exp \left[ \left( \varepsilon _{n}-\mu \right) /k_{%
\text{B}}T\right] +1\right) $ is the Fermi distribution function for the
band $n$, $\varepsilon _{n}$ is the energy of the band $n$, $\mu $ is the
chemical potential, $k_{\text{B}}$ is the Boltzmann constant, and $T$ is the
temperature. We calculate the Drude conductivity numerically for a wide
range of $J/\lambda $\ in Fig.\ref{FigDrude}(b) and (d). To understand the
results analytically, we use a perturbation theory in $J/\lambda $ assuming $%
\left\vert J\right\vert \ll \hbar ^{2}k_{0}/\left( 2\bar{m}\right) $ with $%
k_{0}\equiv \bar{m}\lambda /2\hbar ^{2}$.

The longitudinal Drude conductivity is calculated up to the second order in $%
J/\lambda $ as 
\begin{align}
\sigma _{\text{Drude}}^{x;x}=& \frac{e^{2}\tau }{\hbar ^{2}}\Big[2\pi
\lambda \bar{m}\sqrt{\lambda ^{2}+\frac{2\mu }{\bar{m}}}+\frac{\pi
J_{y}\left( \bar{m}\lambda ^{2}-\mu \right) }{\sqrt{\lambda ^{2}+\frac{2\mu 
}{\bar{m}}}}  \notag \\
& +\frac{5\pi \lambda J_{y}^{2}\left( \bar{m}\lambda ^{2}+3\mu \right) }{%
\left( \lambda ^{2}+\frac{2\mu }{\bar{m}}\right) ^{3/2}}\Big]
\end{align}%
for $-\bar{m}\lambda ^{2}/2<\mu <0$, and 
\begin{eqnarray}
\sigma _{\text{Drude}}^{x;x}=\frac{e^{2}\tau }{\hbar ^{2}}\pi \Big[ &&4\mu +2%
\bar{m}\lambda ^{2}+\bar{m}\lambda J_{y}  \notag \\
&&+\frac{\bar{m}}{2}\left( J_{z}^{2}+3\left( J_{x}^{2}+J_{y}^{2}\right)
\right) \Big]
\end{eqnarray}%
for $\mu >0$. It is shown as a function of the chemical potential $\mu $ in
Fig.\ref{FigDrude}(a) and as a function of $J_{y}$ in Fig.\ref{FigDrude}(b).
Hence, $J_{y}$ is detectable by means of the longitudinal linear Drude
conductivity. The sign change occurs\ in $\sigma _{\text{Drude}}^{x;x}$\
around at $J_{y}\simeq 5\lambda $\ as shown in Fig.\ref{FigDrude}(b).

The transverse Drude conductivity is calculated as%
\begin{equation}
\sigma _{\text{Drude}}^{y;x}=-\frac{e^{2}\tau }{\hbar ^{2}}\pi J_{x}\left( 
\frac{3\mu +\bar{m}\lambda ^{2}}{\sqrt{\lambda ^{2}+\frac{2\mu }{\bar{m}}}}%
+J_{y}\frac{\mu \left( 3\mu +2\bar{m}\lambda ^{2}\right) }{\bar{m}\lambda
\left( \lambda ^{2}+\frac{2\mu }{\bar{m}}\right) ^{3/2}}\right)
\label{SyxDrude}
\end{equation}%
up to the second order in $J/\lambda $ for $-\bar{m}\lambda ^{2}/2\mu <0$,
and%
\begin{equation}
\sigma _{\text{Drude}}^{y;x}=-\frac{e^{2}\tau }{\hbar ^{2}}\bar{m}\pi
J_{x}\left( \lambda -\frac{J_{x}^{2}+J_{y}^{2}}{4\lambda }\right)
\end{equation}%
up to the third order in $J/\lambda $ for $\mu >0$. It is proportional to $%
J_{x}$. It is shown as a function of the chemical potential $\mu $ in Fig.%
\ref{FigDrude}(c) and as a function of $J_{x}$ in Fig.\ref{FigDrude}(d).
Hence, $J_{x}$ is also detectable by means of the transverse linear Drude
conductivity. There is no sign change in $\sigma _{\text{Drude}}^{y;x}$\ as
a function of $J_{x}$\ as shown in Fig.\ref{FigDrude}(d) in contrast to $%
\sigma _{\text{Drude}}^{x;x}$.

We note that the sign change occurs at $\mu =-\bar{m}\lambda ^{2}/3$ in Fig.%
\ref{FigDrude}(c), which is understood analytically in Eq.(\ref{SyxDrude}).

Quantitative differences between the analytic results based on the
perturbation theory and numerical results exist in the vicinity of the band
bottom. It is because we assume two Fermi surfaces surrounding the $\Gamma $%
\ point ($k_{x}=k_{y}=0$) as in Eq.(\ref{kpm}) in analytic calculations.
However, actually, there are two Fermi surfaces, which do not surround the $%
\Gamma $\ point as shown in Fig.\ref{FigBandY}(b). This is the reason for
the discrepancy.

\begin{figure}[t]
\centerline{\includegraphics[width=0.48\textwidth]{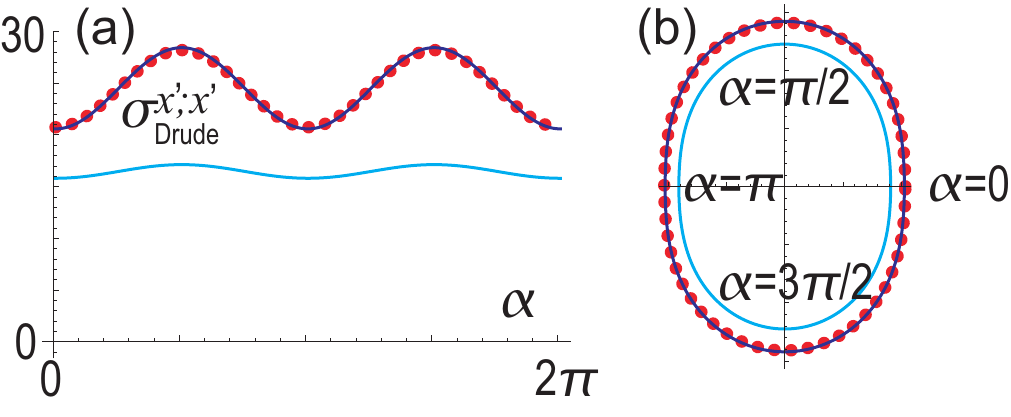}}
\caption{(a) $\protect\sigma _{\text{Drude}}^{x^{\prime };x^{\prime }}$ as a
function of the angle $\protect\alpha $. The vertical axis is $\protect%
\sigma _{\text{Drude}}^{x^{\prime };x^{\prime }}$ in units of $e^{2}\protect%
\tau /\hbar ^{2}$. (b) Corresponding polar plot. The cyan curves represent
analytic results Eq.(\protect\ref{angle}) up to the first order in $J/%
\protect\lambda $, while the dark blue curves represent analytic results Eq.(%
\protect\ref{angle}) up to the second order in $J/\protect\lambda $. Red
dots indicate numerically obtained results. We have set $J=0.5\protect%
\varepsilon _{0}/k_{0}$, $\Theta =\protect\pi /2$, $\Phi =\protect\pi /2$, $%
\hbar ^{2}k_{0}^{2}/\left( 2m\right) =\protect\varepsilon _{0}/2$, $\protect%
\mu =0.2\protect\varepsilon _{0}$ and $\protect\lambda =\protect\varepsilon %
_{0}/k_{0}$.}
\label{FigAngle}
\end{figure}

Next, we study the angle dependence of the Drude conductivity, where the
applied electric field and the lead to read out the conductivity are rotated
by the angle $\alpha $, 
\begin{align}
k_{x}^{\prime }& =k_{x}\cos \alpha -k_{y}\sin \alpha , \\
k_{y}^{\prime }& =k_{x}\sin \alpha +k_{y}\cos \alpha .
\end{align}%
The derivatives are%
\begin{align}
\frac{\partial }{\partial k_{x}^{\prime }}& =\cos \alpha \frac{\partial }{%
\partial k_{x}}-\sin \alpha \frac{\partial }{\partial k_{y}}, \\
\frac{\partial }{\partial k_{y}^{\prime }}& =\sin \alpha \frac{\partial }{%
\partial k_{x}}+\cos \alpha \frac{\partial }{\partial k_{y}}.
\end{align}%
The Drude conductivity along the $x^{\prime }$\ direction is calculated as%
\begin{equation}
\sigma _{\text{Drude}}^{x^{\prime };x^{\prime }}=\sigma _{\text{Drude}%
}^{x;x}\cos ^{2}\alpha -\sigma _{\text{Drude}}^{y;x}\sin 2\alpha +\sigma _{%
\text{Drude}}^{y;y}\sin ^{2}\alpha ,
\end{equation}%
or%
\begin{align}
& \sigma _{\text{Drude}}^{x^{\prime };x^{\prime }}  \notag \\
& =-\frac{e^{2}\tau }{\hbar ^{2}}\pi \lbrack 2\bar{m}\lambda ^{2}+4\mu -\bar{%
m}\lambda \left( J_{x}\sin 2\alpha +J_{y}\left( \cos 2\alpha -2\right)
\right)  \notag \\
& -\frac{\bar{m}}{2}\left( \left( \cos 2\alpha -4\right) \left(
J_{x}^{2}+J_{y}^{2}\right) +\left( \cos 2\alpha -2\right) J_{z}^{2}\right) ]
\label{angle}
\end{align}%
up to the second order in $J/\lambda $ for $\mu >0$. The angle dependence of
the Drude conductivity is shown in Fig.\ref{FigAngle}. It has the $p$-wave
symmetry as shown in Fig.\ref{FigAngle}(b).

\subsection{Anomalous Hall conductivity}

In the following, we consider the case $B\neq 0$ in the Hamiltonian (\ref%
{Zeeman}).

The anomalous Hall conductivity is given by the formula%
\begin{equation}
\sigma _{\text{AHE}}^{y;x}=\frac{e^{2}}{\hbar }\sum_{n}\int d^{2}{k}%
f_{n}\Omega _{n},
\end{equation}%
where the Berry curvature $\Omega _{\pm }$\ is defined by%
\begin{equation}
\Omega _{\pm }\left( \mathbf{k}\right) \equiv \partial _{x}A_{\pm
}^{y}\left( \mathbf{k}\right) -\partial _{y}A_{\pm }^{x}\left( \mathbf{k}%
\right)
\end{equation}%
with the Berry connection%
\begin{equation}
A_{\pm }^{\mu }\left( \mathbf{k}\right) \equiv i\left\langle \psi _{\pm
}\left( \mathbf{k}\right) \right\vert \partial _{k_{\mu }}\left\vert \psi
_{\pm }\left( \mathbf{k}\right) \right\rangle .
\end{equation}%
The Berry curvature $\Omega _{\pm }$\ for the band $\varepsilon _{\pm }$ is
calculated as%
\begin{equation}
\Omega _{\pm }=\pm \frac{B}{4\lambda ^{2}k^{3}}\left( 2\lambda +3J_{x}\sin
2\phi -J_{y}\left( 3\cos 2\phi -1\right) \right)  \label{Omega}
\end{equation}%
up to the first order in $B/\bar{m}\lambda ^{2}$ and $J/\lambda $. The
anomalous Hall conductivity is obtained as%
\begin{equation}
\sigma _{\text{AHE}}^{y;x}=\frac{e^{2}}{\hbar }\frac{B\pi }{\lambda }\left( 
\frac{\sqrt{\lambda ^{2}+\frac{2\mu }{\bar{m}}}}{\mu }-\frac{J_{y}}{m\lambda 
\sqrt{\lambda ^{2}+\frac{2\mu }{\bar{m}}}}\right)  \notag
\end{equation}%
for $\mu <-\left\vert B\right\vert $, 
\begin{equation}
\sigma _{\text{AHE}}^{y;x}=\frac{e^{2}}{\hbar }\frac{B\pi }{\mu }
\end{equation}%
for $\mu >\left\vert B\right\vert $,%
\begin{align}
\sigma _{\text{AHE}}^{y;x}=& \pi \text{sgn}\left[ B\right] -\frac{B}{\sqrt{%
B^{2}+2\bar{m}\lambda ^{2}\left( \mu +\bar{m}\lambda \Lambda \right) }} 
\notag \\
& +J_{y}\frac{\pi B\bar{m}^{2}\lambda ^{2}\Lambda \left( 2\mu +\bar{m}%
\lambda \Lambda \right) }{2\left( 2\mu +\bar{m}\lambda ^{2}\right) \left(
B^{2}+2\bar{m}\lambda ^{2}\left( \mu +\bar{m}\lambda \Lambda \right) \right) 
}
\end{align}%
for $\left\vert \mu \right\vert <\left\vert B\right\vert $ with%
\begin{equation}
\Lambda \equiv \lambda +\sqrt{\lambda ^{2}+\frac{2\mu }{\bar{m}}}.
\end{equation}%
The anomalous Hall conductivity is almost quantized to be $\pm \pi $\ but
slightly deviated from its quantized value. It is because there is a Fermi
surface for $\left\vert \mu \right\vert <\left\vert B\right\vert $.\ Note
that exact quantization occurs in the absence of a Fermi surface between the
Dirac gap.

\begin{figure}[t]
\centerline{\includegraphics[width=0.48\textwidth]{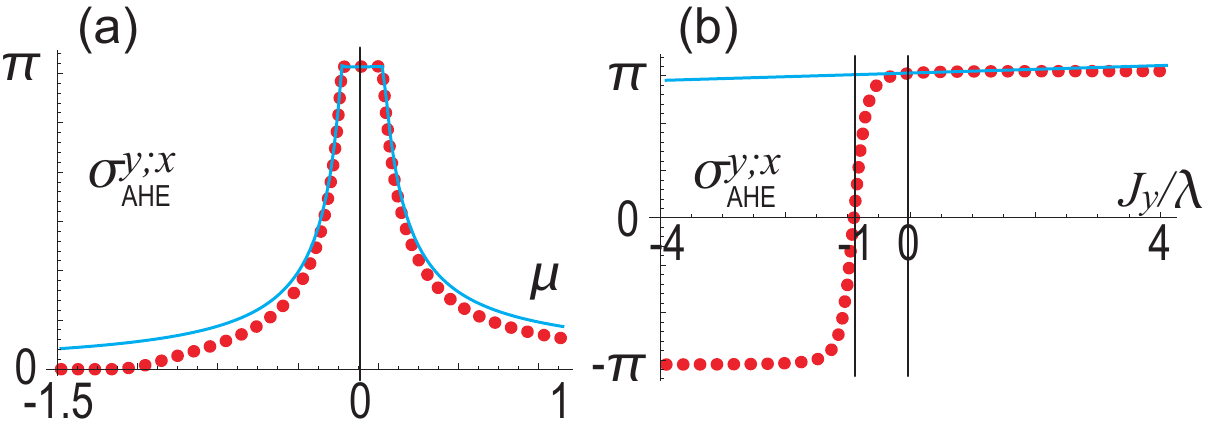}}
\caption{(a) $\protect\mu $ dependence of anomalous Hall conductivities $%
\protect\sigma _{\text{AHE}}^{y;x}$, where\ the spin-splitting vector is
taken along the $y$ axis. The horizontal axis is the chemical potential $%
\protect\mu $ in units of $\protect\varepsilon _{0}$. We have set $J=-0.1%
\protect\varepsilon _{0}/k_{0}$. (b) $J_{y}/\protect\lambda $ dependence of
anomalous Hall conductivities $\protect\sigma _{\text{AHE}}^{y;x}$, where\
the spin-splitting vector is taken along the $y$ axis. The horizontal axis
is $J_{y}/\protect\lambda $. We have set $\protect\mu =0$. Colored dots and
curves are explained in the caption of Fig.\protect\ref{FigDrude}. We have
set $\hbar ^{2}k_{0}^{2}/\left( 2m\right) =\protect\varepsilon _{0}/2$, $%
\protect\lambda =\protect\varepsilon _{0}/k_{0}$, and $B=0.1\protect%
\varepsilon _{0}$. }
\label{FigAHE}
\end{figure}

The anomalous Hall conductivity is shown as a function of $\mu $\ in Fig.\ref%
{FigAHE}(a) as a function of $J_{y}/\lambda $ in Fig.\ref{FigAHE}(b).

\section{Nonlinear conductivity}

\label{Nonlinear}

The second-order nonlinear conductivity $\sigma ^{ab;c}$ is expanded in
terms of the electron relaxation time $\tau $ as\cite{Kaplan}

\begin{equation}
\sigma ^{ab;c}=\sigma _{\text{Metric}}^{ab;c}+\sigma _{\text{Dipole}%
}^{ab;c}+\sigma _{\text{NLDrude}}^{ab;c},
\end{equation}%
where%
\begin{equation}
\sigma _{\text{Metric}}^{ab;c}\propto \tau ^{0},\quad \sigma _{\text{Dipole}%
}^{ab;c}\propto \tau ,\quad \sigma _{\text{NLDrude}}^{ab;c}\propto \tau ^{2}.
\end{equation}

First, only the term $\sigma _{\text{Metric}}^{ab;c}$ survives in the dirty
limit $\tau \rightarrow 0$, which is the intrinsic nonlinear conductivity.
It is the quantum-metric induced nonlinear conductivity given by\cite{Kaplan}
\begin{align}
& \sigma _{\text{Metric}}^{ab;c}  \notag \\
& =-\frac{e^{3}}{\hbar }\sum_{n}\int d^{2}{k}f_{n}\left( 2\frac{\partial
G_{n}^{ab}}{\partial k_{c}}-\frac{1}{2}\left( \frac{\partial G_{n}^{bc}}{%
\partial k_{a}}+\frac{\partial G_{n}^{ac}}{\partial k_{b}}\right) \right) ,
\label{Metric}
\end{align}%
where $G_{n}^{ab}$ is the band--energy normalized quantum metric or the
Berry connection polarizability. It is given as\cite%
{Gao,HLiu,CWang,KamalDas,Kaplan,Teresa}%
\begin{equation}
G_{n}^{ab}=2\text{Re}\sum_{m\neq n}\frac{A_{nm}^{a}\left( \mathbf{k}\right)
A_{mn}^{b}\left( \mathbf{k}\right) }{\varepsilon _{n}\left( \mathbf{k}%
\right) -\varepsilon _{m}\left( \mathbf{k}\right) },  \label{BMetric}
\end{equation}%
with $\varepsilon _{n}$ being the energy of the band $n$, and $A_{nm}^{a}$\
being the interband Berry connection%
\begin{equation}
A_{nm}^{a}\left( \mathbf{k}\right) =i\left\langle \psi _{n}\left( \mathbf{k}%
\right) \right\vert \partial _{k_{a}}\left\vert \psi _{m}\left( \mathbf{k}%
\right) \right\rangle .
\end{equation}

Second, $\sigma _{\text{Dipole}}^{ab;c}$ is the nonlinear transverse
conductivity induced by the Berry curvature dipole (BCD)\cite{Sodeman},%
\begin{equation}
\sigma _{\text{Dipole}}^{ab;c}=-\frac{e^{3}\tau }{\hbar ^{2}}\sum_{n}\int
d^{2}kf_{n}\left( \frac{\partial \Omega _{n}^{bc}}{\partial k_{a}}+\frac{%
\partial \Omega _{n}^{ac}}{\partial k_{b}}\right)
\end{equation}%
with the Berry curvature%
\begin{equation}
\Omega _{n}^{ab}\equiv \partial _{a}A_{nn}^{b}\left( \mathbf{k}\right)
-\partial _{b}A_{nn}^{a}\left( \mathbf{k}\right) .
\end{equation}%
It is an extrinsic nonlinear conductivity, since it vanishes as $\tau
\rightarrow 0$.

Third, $\sigma _{\text{NLDrude}}^{\text{ab;c}}$ is the nonlinear Drude
conductivity\cite{NLDrude},%
\begin{equation}
\sigma _{\text{NLDrude}}^{ab;c}=-\frac{e^{3}\tau ^{2}}{\hbar ^{3}}%
\sum_{n}\int d^{2}kf_{n}\frac{\partial ^{3}\varepsilon _{n}}{\partial
k_{a}\partial k_{b}\partial k_{c}}.  \label{NLDrude}
\end{equation}%
It is also an extrinsic nonlinear conductivity.

\begin{figure}[t]
\centerline{\includegraphics[width=0.48\textwidth]{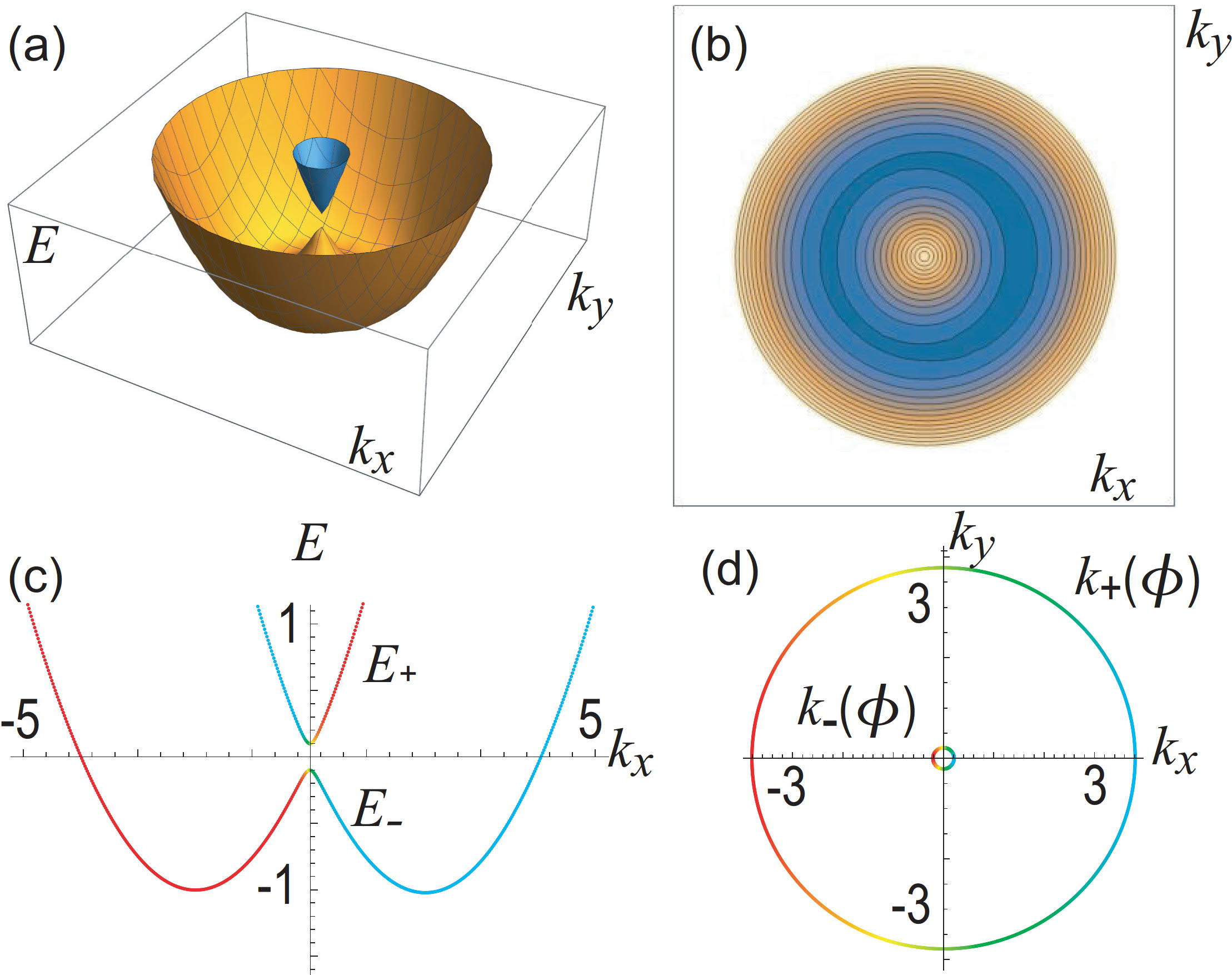}}
\caption{(a) Bird's eye's view of the band structure, where the
spin-splitting vector is taken along the $z$ axis with $\Theta =0$ and $\Phi
=0$. (b) Its contour plot. The color palette is given in Fig.\protect\ref%
{FigBandY}(c). (c) Its cross section along the $k_{x}$ axis. The horizontal
axis is $k_{x}$ in units of $k_{0}$. (d) Fermi surfaces at $\protect\mu =-0.2%
\protect\varepsilon _{0}$. The color palette for (c) and (d) is given in Fig.%
\protect\ref{FigBandY}(f). We have set $\hbar ^{2}k_{0}^{2}/\left( 2m\right)
=\protect\varepsilon _{0}/2$, $\protect\lambda =\protect\varepsilon %
_{0}/k_{0}$, $J=0.5\protect\varepsilon _{0}/k_{0}$, $B=0.1\protect%
\varepsilon _{0}$ and $k_{0}=\bar{m}\protect\lambda /2\hbar ^{2}$. Two Fermi
surfaces are presented as $k_{\pm }\left( \protect\phi \right) $, where $%
k_{x}=k\cos \protect\phi $ and $k_{y}=k\sin \protect\phi $.}
\label{FigBandZ}
\end{figure}

\begin{figure}[t]
\centerline{\includegraphics[width=0.48\textwidth]{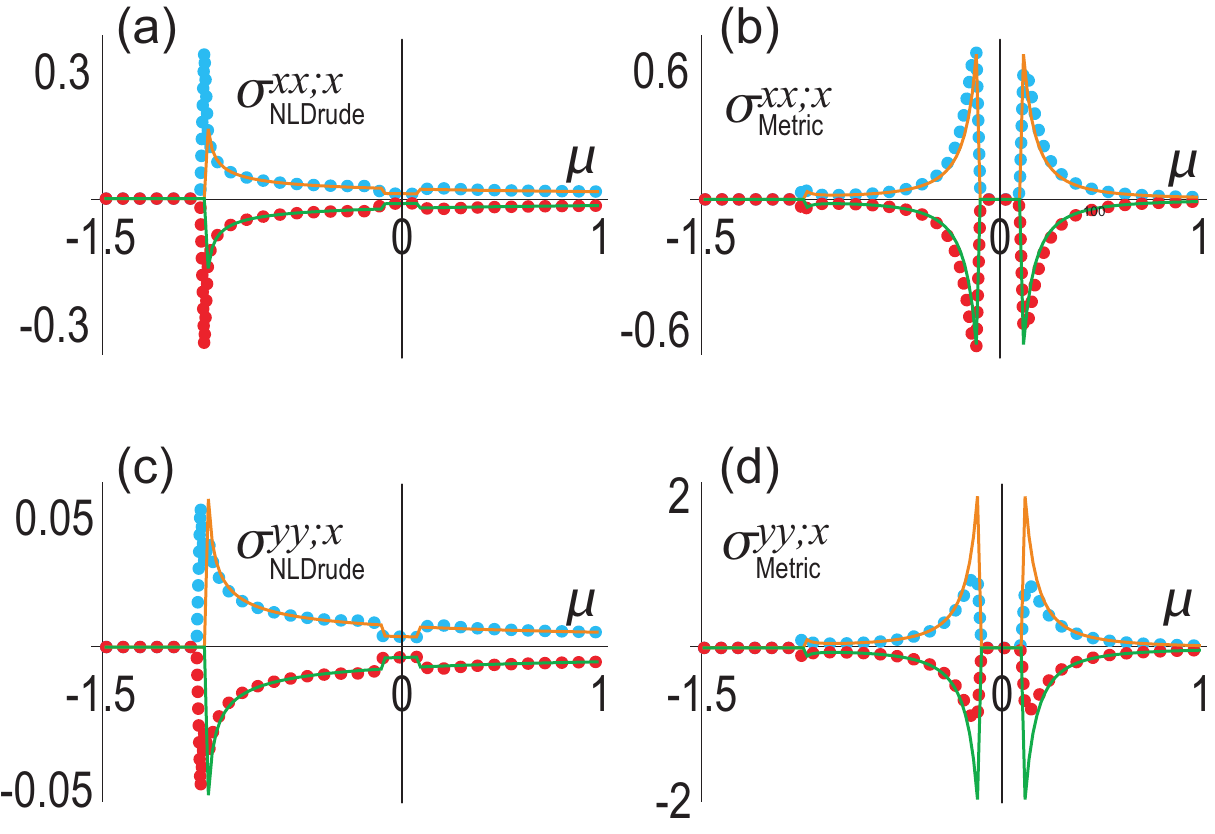}}
\caption{$\protect\mu $ dependence of nonlinear conductivities, where\ the
spin-splitting vector is taken along the $z$ axis. (a) Nonlinear
longitudinal Drude conductivity $\protect\sigma _{\text{Drude}}^{xx;x}$ in
units of $e^{3}\protect\tau ^{2}\protect\varepsilon _{0}/\left( \hbar
^{3}k_{0}\right) $. (b) Nonlinear transverse Drude conductivity $\protect%
\sigma _{\text{NLDrude}}^{yyx;}$ in units of $e^{3}\protect\tau ^{2}\protect%
\varepsilon _{0}/\left( \hbar ^{3}k_{0}\right) $. (c) Quantum-metric induced
longitudinal conductivity $\protect\sigma _{\text{Metric}}^{xx;x}$ in units
of $e^{3}/\left( \hbar \protect\varepsilon _{0}k_{0}\right) $. (d)
Quantum-metric induced transverse conductivity $\protect\sigma _{\text{Metric%
}}^{yy;x}$ in units of $e^{3}/\left( \hbar \protect\varepsilon %
_{0}k_{0}\right) $. The horizontal axis is the chemical potential $\protect%
\mu $ in units of $\protect\varepsilon _{0}$. Colored dots and curves are
explained in the caption of Fig.\protect\ref{FigDrude}. We have set $\hbar
^{2}k_{0}^{2}/\left( 2m\right) =\protect\varepsilon _{0}/2$, $\protect%
\lambda =\protect\varepsilon _{0}/k_{0}$, $J=-0.1\protect\varepsilon %
_{0}/k_{0}$ and $B=0.1\protect\varepsilon _{0}$. }
\label{FigPCon}
\end{figure}

\subsection{Nonlinear longitudinal conductivity}

We proceed to study nonlinear conductivities. The band structure is shown in
Fig.\ref{FigBandZ}, where the spin-splitting vector is taken along the $z$
axis. A Dirac cone at the momentum $k_{x}=k_{y}=0$ has a gap with $2|B|$ as
shown in Fig.\ref{FigBandZ}(a) and (c).\ There are two Fermi surfaces as in
Fig.\ref{FigBandZ}(d) formed by the lower band $\varepsilon _{-}$ for $\mu
<-|B|$, and formed by the lower band $\varepsilon _{-}$ and the upper band $%
\varepsilon _{+}$ for $\mu >|B|$, while there is a Fermi surface formed by
the lower band $E_{-}$ for $|\mu |<|B|$, as shown in Fig.\ref{FigBandZ}(c).

We first study nonlinear longitudinal conductivity. There are two
contributions. One is the nonlinear Drude conductivity\cite{Ideue} and the
other is the quantum-metric induced nonlinear conductivity. We use a
perturbation theory in $J/\lambda $ and $B/\bar{m}\lambda ^{2}$ assuming $%
\left\vert J\right\vert \ll \hbar ^{2}k_{0}/\left( 2\bar{m}\right) $ and $%
\left\vert B\right\vert \ll \hbar ^{2}k_{0}^{2}/\left( 2\bar{m}\right) $.

The nonlinear Drude conductivity at the chemical potential $\mu $ is
analytically calculated based on Eq.(\ref{NLDrude}) as

\begin{equation}
\sigma _{\text{NLDrude}}^{xx;x}=\frac{e^{3}\tau ^{2}}{\hbar ^{3}}\frac{3\pi
\xi \left( \mu \right) }{2\bar{m}\lambda \sqrt{\lambda ^{2}+\frac{2\mu }{%
\bar{m}}}}BJ_{z},
\end{equation}%
where $\xi \left( \mu \right) =1$ for $\left\vert \mu \right\vert
>\left\vert B\right\vert $, and $\xi \left( \mu \right) =1/2$ for $%
\left\vert \mu \right\vert <\left\vert B\right\vert $. It is valid for $\mu
>-\bar{m}\lambda ^{2}/2$. It is proportional to $J_{z}$.

On the other hand, the quantum-metric induced nonlinear conductivity is
analytically obtained based on Eq.(\ref{Metric}) as 
\begin{equation}
\sigma _{\text{Metric}}^{xx;x}=\frac{e^{3}}{\hbar }\frac{5\pi \left( \mu -%
\bar{m}\lambda ^{2}\right) }{8\mu ^{2}\bar{m}^{2}\lambda ^{3}\sqrt{\lambda
^{2}+\frac{2\mu }{\bar{m}}}}BJ_{z}
\end{equation}%
for $\left\vert \mu \right\vert >\left\vert B\right\vert $, and 
\begin{equation}
\sigma _{\text{Metric}}^{xx;x}=\frac{e^{3}}{\hbar }\frac{5\pi \left( \bar{m}%
\lambda \left( \sqrt{\lambda ^{2}+\frac{2\mu }{\bar{m}}}+\lambda \right)
-2\mu \right) }{8\bar{m}^{3}\lambda ^{3}\left( \bar{m}\lambda ^{2}-2\mu
\right) \left( \sqrt{\lambda ^{2}+\frac{2\mu }{\bar{m}}}+\lambda \right) ^{3}%
}BJ_{z}
\end{equation}%
for $\left\vert \mu \right\vert <\left\vert B\right\vert $. It is also
proportional to $J_{z}$. Hence, $J_{z}$ is observable. The leading order of
the nonlinear conductivity is proportional to $BJ$, which means that
magnetization $B$ is necessary to detect $J_{z}$.

\begin{figure}[t]
\centerline{\includegraphics[width=0.48\textwidth]{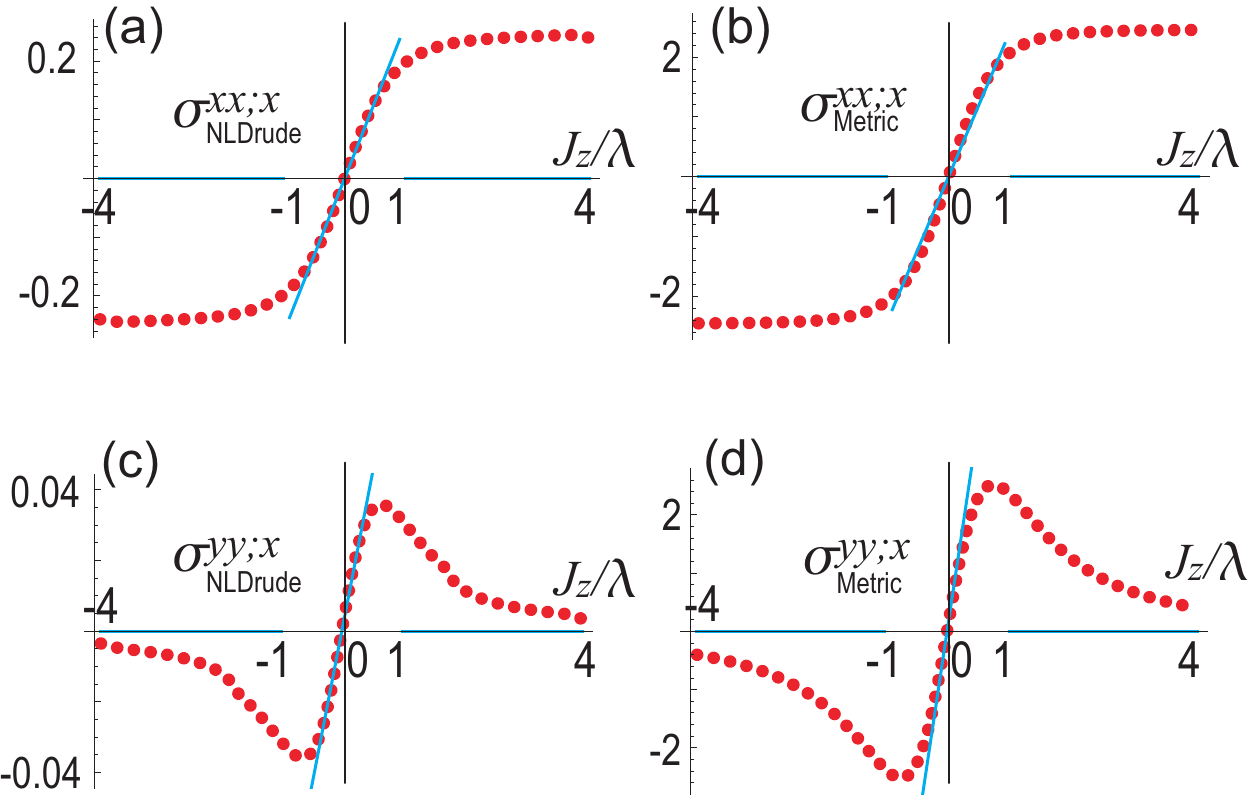}}
\caption{$J/\protect\lambda $ dependence of nonlinear conductivities, where
the spin-splitting vector is taken along the $z$\ axis. (a) Nonlinear
longitudinal Drude conductivity $\protect\sigma _{\text{NLDrude}}^{xx;x}$ in
units of $e^{3}\protect\tau ^{2}E_{0}/\left( \hbar ^{3}k_{0}\right) $. (b)
Quantum-metric induced nonlinear longitudinal conductivity $\protect\sigma _{%
\text{Metric}}^{xx;x}$ in units of $e^{3}/\left( \hbar E_{0}k_{0}\right) $.
(c) Nonlinear transverse Drude conductivity $\protect\sigma _{\text{NLDrude}%
}^{yyx;}$ in units of $e^{3}\protect\tau ^{2}E_{0}/\left( \hbar
^{3}k_{0}\right) $. (d) Quantum-metric induced nonlinear transverse
conductivity $\protect\sigma _{\text{Metric}}^{yy;x}$ in units of $%
e^{3}/\left( \hbar \protect\varepsilon _{0}k_{0}\right) $. The horizontal
axis is the magnetization $J_{z}/\protect\lambda $. Colored dots and curves
are explained in the caption of Fig.\protect\ref{FigDrude}. We have set $%
\hbar ^{2}k_{0}^{2}/\left( 2m\right) =\protect\varepsilon _{0}/2$, $\protect%
\lambda =\protect\varepsilon _{0}/k_{0}$, $B=-0.1\protect\varepsilon %
_{0}/k_{0}$ and $\protect\mu =-0.2\protect\varepsilon _{0}$.}
\label{FigNLJ}
\end{figure}

We show analytical results based on the perturbation theory and numerical
results without using the perturbation theory. The nonlinear conductivity is
shown as a function of the chemical potential $\mu $ in Fig.\ref{FigPCon}.
They agree with each other very well, which assures the validity of the
perturbation theory. The nonlinear Drude conductivity diverges at the band
bottom $\mu =-\bar{m}\lambda ^{2}/2$ and its value becomes half inside of
the bulk gap $\left\vert \mu \right\vert <\left\vert B\right\vert $
comparing with the outside of the bulk gap $\left\vert \mu \right\vert
>\left\vert B\right\vert $. On the other hand, the quantum-metric induced
nonlinear conductivity diverges at the band edge $\left\vert \mu \right\vert
=\left\vert B\right\vert $ and takes tiny value inside of the band gap $%
\left\vert \mu \right\vert <\left\vert B\right\vert $. Therefore, two
contributions are differentiated although the explicit value of $\tau $ is
unknown.

The nonlinear conductivities are shown as a function of $J/\lambda $\ in Fig.%
\ref{FigNLJ}(a) and (b). The $\sigma _{\text{NLDrude}}^{xx;x}$\ and $\sigma
_{\text{Metric}}^{xx;x}$\ approach constant values.

The nonlinear conductivities are shown as a function of $B$ in Fig.\ref%
{FigBCon}(a) and (b). They show nonmonotonic behavior and there is a sudden
jump at $\left\vert B\right\vert =\left\vert \mu \right\vert $. It is
because nonlinear conductivities depend on whether the chemical potential $%
\mu $\ is outside of the band gap $\left\vert B\right\vert $ or inside of
the band gap.

\subsection{Nonlinear transverse conductivity}

Next, we study nonlinear transverse conductivity. There are three
contributions. Nonlinear Drude conductivity, the quantum-metric induced
nonlinear conductivity and the Berry-curvature-dipole induced nonlinear
conductivity.

The nonlinear Drude conductivity at the chemical potential $\mu $ is
analytically calculated based on Eq.(\ref{NLDrude}) as 
\begin{equation}
\sigma _{\text{NLDrude}}^{yy;x}=\frac{e^{3}\tau ^{2}}{\hbar ^{3}}\frac{\pi }{%
2\bar{m}\lambda \sqrt{\lambda ^{2}+\frac{2\mu }{\bar{m}}}}BJ_{z}\xi \left(
\mu \right) .
\end{equation}%
On the other hand, quantum-metric induced nonlinear conductivity is
analytically obtained based on Eq.(\ref{Metric}) as%
\begin{equation}
\sigma _{\text{Metric}}^{yy;x}=\frac{e^{3}}{\hbar }\frac{29\pi \left( \bar{m}%
\lambda ^{2}+\mu \right) }{8\mu ^{2}\bar{m}^{2}\lambda ^{3}\sqrt{\lambda
^{2}+\frac{2\mu }{\bar{m}}}}BJ_{z}
\end{equation}%
for $\left\vert \mu \right\vert >\left\vert B\right\vert $, and%
\begin{equation}
\sigma _{\text{Metric}}^{yy;x}=\frac{e^{3}}{\hbar }\frac{29\pi \left( \bar{m}%
\lambda \left( \sqrt{\lambda ^{2}+\frac{2\mu }{\bar{m}}}+\lambda \right)
-2\mu \right) }{16m^{3}\lambda ^{3}\left( \bar{m}\lambda ^{2}-2\mu \right)
\left( \sqrt{\lambda ^{2}+\frac{2\mu }{\bar{m}}}+\lambda \right) ^{3}}BJ_{z}
\end{equation}%
for $\left\vert \mu \right\vert <\left\vert B\right\vert $. Hence, we can
also observe $J_{z}$ by measuring the nonlinear transverse conductivity. The
behavior of the nonlinear transverse conductivity as a function of $\mu $ is
similar to that of the nonlinear longitudinal conductivity.

The nonlinear conductivities are shown as a function of $J_{z}$\ in Fig.\ref%
{FigNLJ}(c) and (d). The $\sigma _{\text{NLDrude}}^{yy;x}$\ and $\sigma _{%
\text{Metric}}^{yy;x}$\ approach constant values. They are linear in $J_{z}$%
\ but decreases and approach zero for large $J_{z}$. However, there is no
sign change in them.

\begin{figure}[t]
\centerline{\includegraphics[width=0.48\textwidth]{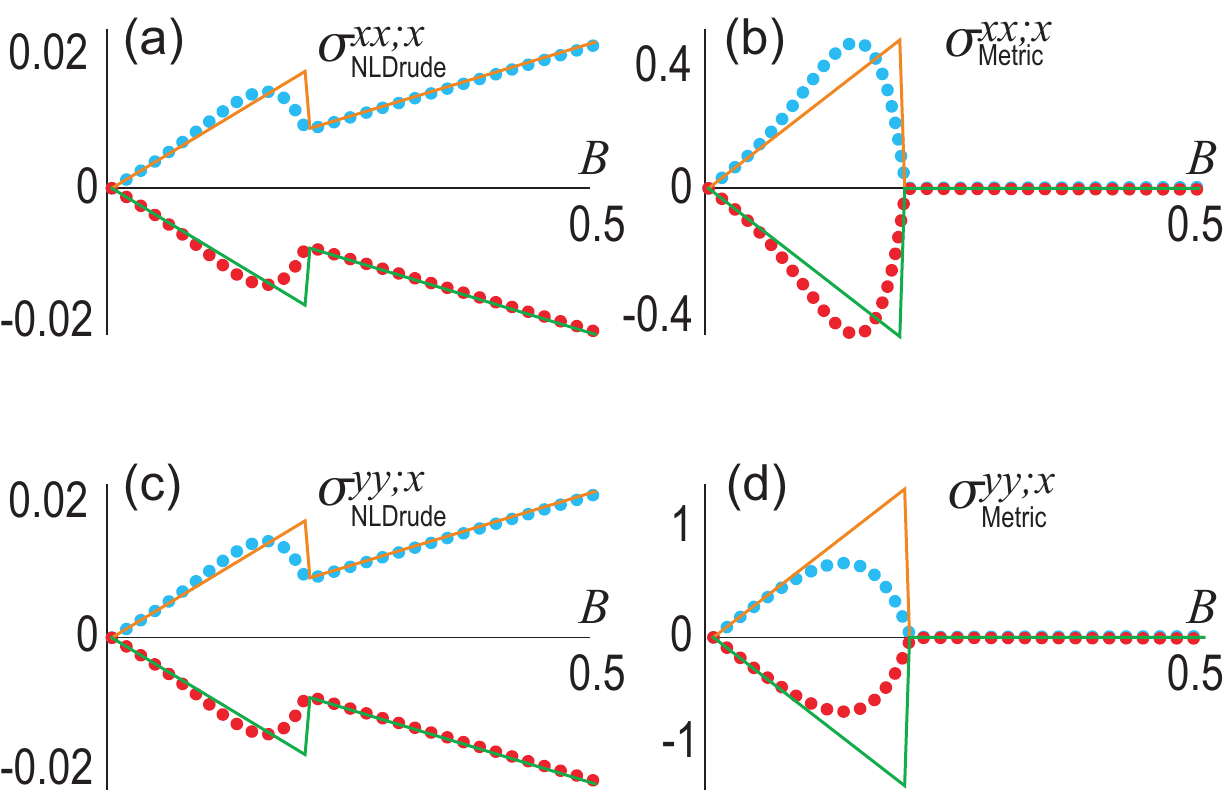}}
\caption{$B$ dependence of nonlinear conductivities, where the
spin-splitting vector is taken along the $z$\ axis. (a) Nonlinear
longitudinal Drude conductivity $\protect\sigma _{\text{NLDrude}}^{xx;x}$ in
units of $e^{3}\protect\tau ^{2}E_{0}/\left( \hbar ^{3}k_{0}\right) $. (b)
Quantum-metric induced nonlinear longitudinal conductivity $\protect\sigma _{%
\text{Metric}}^{xx;x}$ in units of $e^{3}/\left( \hbar E_{0}k_{0}\right) $.
(c) Nonlinear transverse Drude conductivity $\protect\sigma _{\text{NLDrude}%
}^{yyx;}$ in units of $e^{3}\protect\tau ^{2}E_{0}/\left( \hbar
^{3}k_{0}\right) $. (d) Quantum-metric induced nonlinear transverse
conductivity $\protect\sigma _{\text{Metric}}^{yy;x}$ in units of $%
e^{3}/\left( \hbar \protect\varepsilon _{0}k_{0}\right) $. The horizontal
axis is the magnetization $B$ in units of $\protect\varepsilon _{0}$.
Colored dots and curves are explained in the caption of Fig.\protect\ref%
{FigDrude}. We have set $\hbar ^{2}k_{0}^{2}/\left( 2m\right) =\protect%
\varepsilon _{0}/2$, $\protect\lambda =\protect\varepsilon _{0}/k_{0}$, $%
J=-0.1\protect\varepsilon _{0}/k_{0}$ and $\protect\mu =-0.2\protect%
\varepsilon _{0}$.}
\label{FigBCon}
\end{figure}

The nonlinear conductivities are shown as a function of $B$ in Fig.\ref%
{FigBCon}(c) and (d). They show nonmonotonic behavior and there is a sudden
jump at $\left\vert B\right\vert =\left\vert \mu \right\vert $\ as in the
case of the nonlinear longitudinal conductivity. Quantitative differences
exist in the vicinity of the gap $B=\mu $\ in Fig.\ref{FigBCon}, because we
made a perturbation with respect to $B$, which is invalid around the gap.

\subsection{Berry curvature dipole}

The Berry curvature dipole induced nonlinear conductivity is given by%
\begin{equation}
\sigma _{\text{Dipole}}^{xx;y}=-2\frac{e^{3}\tau }{\hbar ^{2}}\sum_{n}\int
d^{2}kf_{n}\frac{\partial \Omega _{n}^{yx}}{\partial k_{x}}.
\end{equation}%
By differentiating Eq.(\ref{Omega}), we have%
\begin{align}
\frac{\partial \Omega _{n}^{yx}}{\partial k_{x}}& =-\frac{3Bk\lambda
^{4}k_{x}}{2\left( k^{2}\lambda ^{2}+B^{2}\right) ^{5/2}}  \notag \\
& -J_{z}\frac{3B^{2}\lambda ^{2}\left( 2B^{2}-\lambda ^{2}\left(
3k^{2}+5\left( k_{x}^{2}-k_{y}^{2}\right) \right) \right) }{4\left(
k^{2}\lambda ^{2}+B^{2}\right) ^{7/2}},
\end{align}%
up to the first order in $J/\lambda $, when the spin-splitting vector is
along the $z$ direction.

\begin{figure}[t]
\centerline{\includegraphics[width=0.48\textwidth]{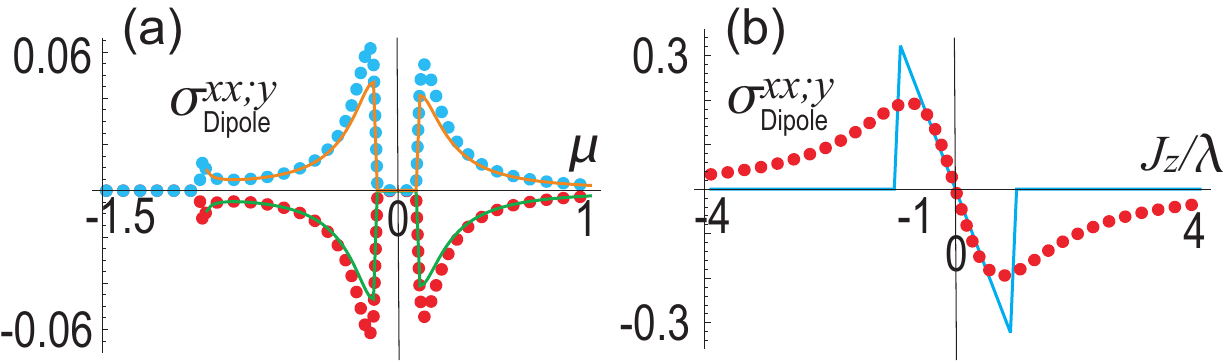}}
\caption{(a) $\protect\mu $ dependence of Berry curvature dipole induced
nonlinear conductivities $\protect\sigma _{\text{Dipole}}^{xx;y}$, where\
the spin-splitting vector is taken along the $z$ axis. The horizontal axis
is the chemical potential $\protect\mu $ in units of $\protect\varepsilon %
_{0}$. We have set $J=-0.1\protect\varepsilon _{0}/k_{0}$. (b) $J_{z}/%
\protect\lambda $ dependence of Berry curvature dipole induced nonlinear
conductivities $\protect\sigma _{\text{Dipole}}^{xx;y}$, where\ the
spin-splitting vector is taken along the $z$ axis. The horizontal axis is $%
J_{z}/\protect\lambda $. We have set $\protect\mu =0$. Colored dots and
curves are explained in the caption of Fig.\protect\ref{FigDrude}. We have
set $\hbar ^{2}k_{0}^{2}/\left( 2m\right) =\protect\varepsilon _{0}/2$, $%
\protect\lambda =\protect\varepsilon _{0}/k_{0}$, and $B=0.1\protect%
\varepsilon _{0}$. }
\label{FigDipole}
\end{figure}

The Berry curvature dipole induced nonlinear conductivity is analytically
obtained as%
\begin{align}
\sigma _{\text{Dipole}}^{xx;y}=& \frac{J_{z}\bar{m}\pi }{2\left( B^{2}+2\bar{%
m}\lambda ^{2}\left( \mu +\bar{m}\lambda \Lambda _{\pm }\right) \right)
^{5/2}}  \notag \\
& \times \Big[\sum_{\pm }\frac{5\bar{m}^{3}\lambda ^{5}\Lambda _{\pm }^{4}}{%
\Lambda _{0}}+\sum_{\pm }6B^{2}\lambda ^{2}\left( \mu +\bar{m}\lambda
\Lambda _{\pm }\right)  \notag \\
& \hspace{3mm}-\frac{\lambda ^{3}\Lambda _{\pm }\left( 3B^{2}+10\bar{m}%
\lambda ^{2}\left( \mu +\bar{m}\lambda \Lambda _{\pm }\right) \right) }{%
\Lambda _{0}}\Big]
\end{align}%
for $\left\vert \mu \right\vert >\left\vert B\right\vert $ and 
\begin{equation}
\sigma _{\text{Dipole}}^{xx;y}=0
\end{equation}%
for $\left\vert \mu \right\vert <\left\vert B\right\vert $.%
\begin{align}
\Lambda _{\pm }& \equiv \lambda \pm \sqrt{\lambda ^{2}+\frac{2\mu }{\bar{m}}}%
, \\
\Lambda _{0}& \equiv \sqrt{\lambda ^{2}+\frac{2\mu }{\bar{m}}}.
\end{align}%
The anomalous Hall conductivity is shown as a function of $\mu $\ in Fig.\ref%
{FigDipole}(a) as a function of $J$ in Fig.\ref{FigDipole}(b).

\section{Discussion}

\label{Discussion}

$p$-wave magnets have zero-net magnetization, which may lead to a
high-density and ultra-fast memory by using the spin-splitting vector as
the\ Ising variable as in the case of antiferromagnets and altermagnets. We
have found that $J_{x}$ and $J_{y}$ are detectable by the linear Drude
conductivity without using magnetization. It is contrasted to the $d$-wave
altermagnet, where the N\'{e}el vector is not detectable by the linear Drude
conductivity and the measurement of the nonlinear conductivity is necessary%
\cite{EzawaAlter}. On the other hand, $J_{z}$ is detectable by measuring the
nonlinear conductivity with the aid of magnetization.

We have analyzed the Hamiltonian (\ref{TotalHamil}) numerically to obtain
results for a wide range of $J/\lambda $, and perturbatively to obtain
analytical results for the strong Rashba regime $\left\vert \lambda
/J\right\vert \gg 1$. It is numerically checked that the sign of the
conductivities are identical except for the longitudinal Drude conductivity.
Hence, $J\left( \mathbf{s}\cdot \mathbf{\sigma }\right) $ is detectable in a
wider range of $\left\vert J/\lambda \right\vert $. In the current
experimental stage, it is difficult to identify whether strong or weak
Rashba limit is realized in a particular sample. We may expect the weak
Rashba interaction if the $p$-wave magnet is metallic, while we may expect
the strong Rashba interaction if the $p$-wave magnet is insulating and the
free fermions exist at the interface.

We make a comment on the spin-orbit interaction. We have determined the
direction of the spin-splitting vector by comparing it with the direction of
the Rashba term by purely electrical means. Indeed, the energy is
independent of the direction of the spin-splitting vector when there is no
spin-orbit interaction, where the linear and nonlinear Drude conductivities
are independent of the direction of the spin-splitting vector. Namely, the
spin-orbit interaction is necessary to readout the direction of the
spin-splitting vector.

The author is very much grateful to S. Okumura and M. Hirschberger for
helpful discussions on the subject. This work is supported by CREST, JST
(Grants No. JPMJCR20T2) and Grants-in-Aid for Scientific Research from MEXT
KAKENHI (Grant No. 23H00171).

\end{document}